\documentclass{article}
\usepackage{graphicx}
\usepackage[centertags]{amsmath}
\usepackage{amsfonts}
\usepackage{amssymb}
\usepackage{amsthm}

\title{Geometric Nonlinearities in Field Theory, Condensed Matter and Analytical Mechanics}

\author{J. J. S\l awianowski\\
Institute of Fundamental Technological Research,\\
Polish Academy of Sciences,\\
$5^{\rm B}$, Pawi\'{n}skiego str., 02-106 Warsaw, Poland\\
e-mail: jslawian@ippt.gov.pl}

\begin{document}

\maketitle
\begin{abstract}
There are two very important subjects in physics: Symmetry of dynamical models and nonlinearity. All really fundamental models are invariant under some particular symmetry groups. There is also no true physics, no our Universe and life at all, without nonlinearity. Particularly interesting are essential, non-perturbative nonlinearities which are not described by correction terms imposed on some well-defined linear background. Our idea in this paper is that there exists some mysterious, not yet understood link between essential, physically relevant nonlinearity and dynamical symmetry, first of all, large symmetry groups. In some sense the problem is known even in soliton theory, where the essential nonlinearity is often accompanied by the infinite system of integrals of motion, thus, by infinite-dimensional symmetry groups. Here we discuss some more familiar problems from the realm of field theory, condensed matter physics, and analytical mechanics, where the link between essential nonlinearity and high symmetry is obvious, even if not yet fully understood. \end{abstract}

\section{Symmetry versus nonlinearity in metrical and tetrad gravitation. Comparison with mechanical toy models}

Something close to the Anthropic Principle and similar ideas: 
\begin{center}
\emph{There is no Our Cosmos and no Life without Nonlinearity. }
\end{center}
In various aspects, quite trivial, every-day-life ones, and very fundamental structural problems. Concerning the every-day life, e.g., there is no thermal
expansion of bodies without nonlinearity expressed by the non-symmetric
shape of the diagram of the interparticle potential energy as a function
of distance. Biological and ecological systems are based on the limit
cycles, impossible without nonlinearity. Similarly, without nonlinearity, 
at least in some background, it is impossible to reconcile the field equations and equations of motion, e.g., in Maxwell electrodynamics. Without nonlinearity there is no stochastization,  no equipartition of energy, therefore, no thermodynamics. It is instructive
to think about some crazy model of condensed matter as a system of mutually
coupled harmonic oscillators, based, e.g., on the isotropic Lagrangians
of the form:
\begin{equation}
L=\frac{1}{2}\underset{A}{\sum}m_{A}\frac{\mathrm{d}x^{i}{}_{A}}{\mathrm{d}t}
\frac{\mathrm{d}x^{j}{}_{A}}{\mathrm{d}t}g_{ij}-\frac{1}{2}\underset{A\neq B}{\sum}\varkappa_{AB}\left(x^{i}{}_{A}-x^{i}{}_{B}\right)
\left(x^{j}{}_{A}-x^{j}{}_{B}\right)g_{ij}, \label{eq:1}
\end{equation}
where $g$ denotes the metric tensor, $\varkappa_{AB}=\varkappa_{BA}$
are the elastic constants, and $m_{A}$ are the particle masses. The corresponding
equations of motion have the form:
\begin{equation}
m_{A}\frac{\mathrm{d}^{2}}{\mathrm{d}t^{2}}x^{i}{}_{A}=
-\underset{B}{\sum}\varkappa_{AB}
\left(x^{i}{}_{A}-x^{i}{}_{B}\right). \label{eq:2}
\end{equation}

Let us notice that the metric tensor does not enter (\ref{eq:2})
at all, although it is explicitly present in (\ref{eq:1}). This is
one of its ambiguous roles in equations of physics. But, never mind, 
the point is that the model (\ref{eq:1}), (\ref{eq:2}) does prevent the decay of the system, but it does not prevent its collapse. The only anti-collapse mean of (\ref{eq:1}), (\ref{eq:2}) is the centrifugal barrier. But this is physically non-sufficient, and the true anti-collapse repulsive potentials must be positively-singular
at coincidences of particles; the corresponding forces will be certainly
non-harmonic and the model essentially nonlinear. As harmonic models
always split into mutually non-interacting one-dimensional normal
modes, in quantum field theory, and classical field theory as its
kindergarten, one accepts the view that the true interaction is encoded
within the anharmonic sector. From this point of view the harmonic
models are ``non-interacting'', although there exist
some ``springs'' between non-normal
modes. Finally, let us repeat the immortal, in any case not yet solved, 
problem of quantum decoherence. There is an ``infinity'' of ideas about it, one of them is some fundamental nonlinearity hidden somewhere beyond the usual pragmatic framework of linear quantum mechanics. 

Nonlinearity is physically desirable, just unavoidable, but, at the
same time, linear models are in principle explicitly treatable. This
motivates a kind of compromise often dealt with in practice. Namely, 
one considers some linear background model with additionally extra
imposed some nonlinear perturbations. This perturbation is often considered
as ``small'', or to be more precise, 
it is controlled by some coupling parameter. The vanishing value of
this parameter corresponds to the background linear model. To solve
nonlinear problems, one employs certain perturbation techniques, expansions
with respect to the ``small'' parameter, 
and the search of solutions in terms of asymptotic series (by collecting
coefficients at the same power of the parameter). Of course, such a procedure
is always more or less ``tricky'', 
certainly non-reliable. One never knows a priori if the underlying
linear background is structurally stable under perturbations. And, 
what is more important, there exist fundamental theories and models
which are essentially nonlinear. They are nonlinear from the very
beginning and there is no natural splitting into linear background
and nonlinear correction term. Let us mention general relativity, 
'tHooft-Polyakov-Kleinert strings, Born-Infeld electrodynamics and
its generalizations, Euler equation for ideal fluids are profound
examples in fundamental field theory and condensed matter physics;
incidentally, the two disciplines are not sharply distinct, and the
border between them is rather diffused \cite{BPT2009,Erin62,Erin68,16,PB2008,PBGT2008,26,jjs_07}. 

In fundamental theories it happens very often that the tensorial structure
of considered objects just canonically induces certain coupling schemes
and certain canonical nonlinearities. It is just the case with the
mentioned examples, where one is faced with the peculiar convolution
of two things: the essential non-perturbative nonlinearity and the
huge symmetry groups. This convolution is in no way accidental and may
be heuristically explained within the framework of variational theories. 
This is not the essential restriction, because usually, dissipative
models preassume certain self-adjoint background, and besides, with
certain modifications, the very idea works for them as well. Simply, 
it is particularly easy to understand them using Lagrangian concepts. 
Namely, from the geometric point of view Lagrangian is a scalar W-density of weight one, built of dynamical variables, i.e., ``fields'', 
and their derivatives with respect to independent variables, let us
say, ``space-time'' coordinates. 
In fundamental theories one deals in principle with first-order derivatives, 
however with certain delicate points concerning general relativity. 
But to construct scalar densities or scalars from ``fields'', 
one needs usually certain ``tools'', 
which enable one to define invariant derivatives with respect to ``space-time''
coordinates, and to contract tensorial spatio-temporal or internal
indices. In specially-relativistic or Galilean physics those tools
are usually some metric tensors and their by-products like affine
connections, volume forms, etc. They are absolute objects of the theory. 
When they are kept fixed, symmetries of the theory are rather poor, 
because they must respect, preserve those objects. In linear theories
the metric tensors enter Lagrangians via coefficients of quadratic
forms built of dynamical quantities, e.g., in kinetic energy, in kinetical
terms of field Lagrangians, etc. \cite{jjs_07}. It enters also through covariant
derivatives of fields, integration element, etc. When kept fixed as
an absolute, controlling object, it restricts the symmetry group to
the finite-dimensional isometry group of $g$. But it was just the
general covariance idea of Hilbert that no absolute objects may exist
in really fundamental theory \cite{jjs_07}. If so, $g$ must be included into physical degrees of freedom and then the symmetry group of the theory becomes
just ${\rm Diff}\; M$, the group of all diffeomorphisms of the space-time
manifold $M$. This is a huge, infinite-dimensional group. And automatically
the theory becomes essentially nonlinear, without any linear background
to be perturbed. And this is a rule: in linear variational theories
the quadratic forms underlying Lagrangians, automatically restrict
the symmetry group to some (pseudo-)Euclidean group. To escape
this restriction, one must include the quadratic form itself into
degrees of freedom, and this self-interaction brings about some essential
nonlinearity \cite{26,jjs_07}. Non-Abelian gauge theories provide another example of the relationship between essential nonlinearity and symmetry groups. The inherent nonlinearity and self-interaction
of gauge fields (the ``radiating radiation'', so to speak) is exactly
due to their symmetry group. Let us also mention about solitons, where
one observes the very peculiar coincidence of the essential nonlinearity
and the rich groups of hidden symmetries (hierarchy of constants of
motion, the total intergrability). Generalized Born-Infeld-type models
(including 'tHooft-Polyakov-Kleinert) offer some very interesting
mechanism of essential nonlinearity, apparently without a direct link to symmetry \cite{BB79,BI34,20,jjs_07}. However, the more detailed analysis shows that some important symmetries are also intimately connected with them. 

It is very instructive to review the structure of nonlinearities quoted
above, with the special stress on their geometric background, first
of all, but not only, on symmetry groups. Certain common features
of field theory, mechanics of continua, condensed matter theory, and
analytical mechanics are then exhibited and the borders between them
diffuse in a sense. 

In general relativity, the Hilbert Lagrangian of the
metric field $g$ on the space-time manifold $M$ is given by \cite{15}
\begin{equation}
\mathcal{L}_{\rm H}\left[g\right]=\mathcal{L}_{\rm H}\left(g, \partial g, \partial^{2}g\right)=-\frac{1}{2\varkappa}\mathcal{R}[g]
\sqrt{\left|g\right|}\label{eq:3}
\end{equation}
with the obvious meaning of symbols: $\varkappa$ is proportional
to the gravitation constant (the proportionality factor depends on
the system of units), $\mathcal{R}[g]$ is the curvature scalar built
of $g$, and $\left|g\right|$ is an abbreviation for the absolute
value of $\det[g_{\mu\nu}]$ in a given coordinate system. Geometrically
$\left|g\right|$ is a scalar $W$-density of weight two, and because
of this, $\mathcal{L}_{\rm H}[g]$ is a scalar $W$-density of weight
one, just as any correctly defined Lagrangian should be. Sometimes
one modifies (\ref{eq:3}) by adding the cosmological term 
\begin{equation}
\mathcal{L}_{\rm cosm}[g]=\Lambda\sqrt{\left|g\right|}, \label{eq:4}
\end{equation}
$\Lambda$ is here a constant parameter usually referred to as cosmological
constant. 

Some comments are necessary here. Namely, Lagrangian (\ref{eq:3})
depends on second derivatives, but the corresponding variational principle
is essentially first-order one. The point is that $\mathcal{L}_{\rm H}$
depends on second derivatives quasilinearly, i.e., linearly with coefficients
depending algebraically on $g$, but not on its first derivatives. 
The second derivatives in (\ref{eq:3}) may be absorbed into a total
divergence term and removed from the action functional, 
\begin{equation}
\mathcal{L}_{\rm H}[g]=
G_{\rm H}[g]
\sqrt{\left|g\right|}+\textrm{``Div''}=
G_{\rm H}(g,\partial g)
\sqrt{\left|g\right|}+\textrm{``Div''}. 
\label{eq:5}
\end{equation}
The first-order Lagrangian $G_{\rm H}\sqrt{\left|g\right|}$ is ``non-aesthetic''
in that it is not a scalar density of weight one, instead, it is a
strange ``object'' which transforms
under the change of coordinates as a density modulo some additive correction
by a total divergence. Nevertheless it works. Hilbert, led by its mathematical
intuition, guessed (\ref{eq:3}) immediately as the only geometrically
correct possibility (up to the ``cosmological''
term). Unlike this, the back-breaking attempts by Einstein were based
on rather qualitative physical ideas and full of mistakes, sometimes
rather funny ones. 

The structurally dominant term of (\ref{eq:3}), (\ref{eq:5}) has
the form (modulo constant factor):
\begin{equation}
g^{\mu\nu}g^{\alpha\gamma}g^{\beta\delta}\left(\partial_{\mu}
g_{\alpha\beta}\right)\left(\partial_{\nu}g_{\gamma\delta}\right)
\sqrt{\left|g\right|}. \label{eq:6}
\end{equation}
Obviously, without the next terms, this is a completely non-tensorial
expression \cite{Kob-Nom63,Ster64}, but it just focuses and visualizes the very essence of nonlinear self-interaction of $g$. 

Lagrangian of matter fields, denoted symbolically by $\Psi$, is given by 
\begin{equation}
\mathcal{L}_{\rm matt}[g, \Psi]=\mathcal{L}_{\rm matt}\left(g, \partial g;\Psi, \partial\Psi\right). \label{eq:7}
\end{equation}
The metric $g$ is used here for contracting the tensorial indices
of $\Psi$, and its first derivatives $\partial g$ occur in the Levi-Civita
affine connection used for tensorially invariant differentiation of
$\Psi$. The total Lagrangian 
\begin{equation}
\mathcal{L}_{\rm tot}[g, \Psi]:=\mathcal{L}_{\rm H}[g]+
\mathcal{L}_{\rm matt}[g, \Psi]\label{eq:8}
\end{equation}
is invariant under the huge infinite-dimensional group ${\rm Diff}\; M$
of all diffeomorphisms of $M$ onto itself, 
\begin{equation}
\mathcal{L}_{\rm tot}[\varphi_{*}g, \varphi_{*}\Psi]=
\varphi_{\ast}\mathcal{L}_{\rm tot}[g, \Psi]\label{eq:9}
\end{equation}
for any $\varphi\in {\rm Diff}\; M$. This concerns separately both terms, 
and obviously the action functional, just as its both terms separately, 
is invariant in the sense: 
\begin{equation}
I[g, \Psi|\Omega]=I[\varphi_{*}g, \varphi_{*}\Psi|\varphi(\Omega)]. \label{eq:10}
\end{equation}
Obviously, the action over the $\Omega$-domain is given by 
\begin{equation}
I[g, \Psi|\Omega]=\underset{\Omega}{\int}\mathcal{L}[g;\Psi]
\mathrm{d}_{4}x;\label{eq:11}
\end{equation}
it is a well-defined scalar because $\mathcal{L}$ is a scalar density
of weight one. The gravitational Hilbert and matter actions $I_{\rm H}$, 
$I_{\rm matt}$ are defined separately in the same way. If $\mathcal{L}_{\rm H}$
is replaced by $G_{\rm H}\sqrt{|g|}$, then the invariance (\ref{eq:10})
is replaced by the invariance modulo some additive term depending
only on the values of fields on the boundary $\partial\Omega$. Obviously, 
this does not affect the invariance of the Euler-Lagrange field equations. 

Let us remind that the field equations have the form: 
\begin{eqnarray}
R_{\mu\nu}-\frac{1}{2}Rg_{\mu\nu} & = & \varkappa T_{\mu\nu}, \nonumber\\
\label{eq:12}\\
\frac{\partial\mathcal{L}_{\rm matt}}{\partial\Psi^{A}}-
\frac{D}{Dx^{\mu}}\frac{\partial\mathcal{L}}{\partial\Psi^{A}{}, _{\mu}} & = & 
0, \nonumber 
\end{eqnarray}
where $R_{\mu\nu}=R^{\alpha}{}_{\mu\alpha\nu}$ is the Ricci tensor, 
$R=g^{\mu\nu}R_{\mu\nu}$ is the curvature scalar, and $T_{\mu\nu}$
is the metrical (thus, symmetric) energy-momentum tensor of matter, 
\begin{equation}
T_{\mu\nu}=-\frac{2}{\sqrt{|g|}}\frac{\delta I_{\rm matt}}{\delta g^{\mu\nu}}, \label{eq:13}
\end{equation}
where, obviously, one must carefully distinguish between covariant
and contravariant components of tensors. In particular, on the right-hand
side of (\ref{eq:13}) one performs the variational procedure with
respect to the contravariant inverse of $g$. 

Without matter, i.e., in empty space-time, field equations reduce to 
\begin{equation}
R_{\mu\nu}=0, \label{eq:14}
\end{equation}
and, in analogy to (\ref{eq:6}), the first, leading term has the
d'Alembert structure, 
\begin{equation}
g^{\mu\nu}\partial_{\mu}\partial_{\nu}g_{\alpha\beta}+\ldots=0. \label{eq:15}
\end{equation}
Of course, this gives a correct insight into the dynamical structure
of field equations, but one must remember that the ``d'Alembert'' term is not to be meant literally, 
because it has no well-defined tensorial structure. 

It is interesting to mention a finite-dimensional counterpart of this
framework, one within the domain of Hamiltonian dynamics \cite{Abr-Mars78,Arn78,Gold50,Ster64,Syn60}. Namely, let
us imagine some physics the area of which is not a general differential
manifold, but rather some affine space $M$ with the linear space
of translations $V$. Instead of relativistic four-dimensional metric
of the normal-hyperbolic signature, we have an Euclidean metric (positive
one) in the usual space. Assume this metric to be a dynamical object, 
and its ``kinetic energy'' to be
an expression of the form:
\begin{equation}
T[g]=\frac{I}{2}g^{ik}g^{jl}\frac{\mathrm{d}g_{ij}}{\mathrm{d}t}
\frac{\mathrm{d}g_{kl}}{\mathrm{d}t}+
\frac{K}{2}g^{ij}g^{kl}\frac{\mathrm{d}g_{ij}}{\mathrm{d}t}
\frac{\mathrm{d}g_{kl}}{\mathrm{d}t}, \label{eq:16}
\end{equation}
$K, I$ being constants. 

In analogy to Hamiltonian systems on groups, 
this may be written as:
\begin{equation}
T[g]=\frac{I}{2}{\rm Tr}\left(\Omega_{l}{}^{2}\right)+\frac{K}{2}
\left({\rm Tr}\; \Omega_{l}\right)^{2}=\frac{I}{2}{\rm Tr}\left(\Omega_{r}{}^{2}\right)+
\frac{K}{2}\left({\rm Tr}\; \Omega_{r}\right)^{2}, \label{eq:17}
\end{equation}
where the quantities $\Omega_{l}\in V\otimes V^{*}\simeq {\rm L}(V)$, $\Omega_{r}\in V^{*}\otimes V$ are defined as follows:
\begin{equation}
\Omega_{l}{}^{a}{}_{b}:=g^{ac}\frac{\mathrm{d}g_{cb}}{\mathrm{d}t}, \qquad
\Omega_{r}{}_{a}{}^{b}=\frac{\mathrm{d}g_{ac}}{\mathrm{d}t}g^{cb}. \label{eq:18}
\end{equation}
Of course, in spite of certain formal similarities, one should be aware
of the difference between this model and that of Hamiltonian systems
on groups \cite{Abr-Mars78,Arn78,Bog85,Mars-Rat94,Mars-Rat99}. 

It is easily seen that (\ref{eq:16}), (\ref{eq:17}) is a finite-dimensional
toy model of (\ref{eq:3}), (\ref{eq:5}), (\ref{eq:6}). The self-interaction
structure of $g$ and its characteristic, non-perturbative nonlinearity
is in principle like in the Hilbert principle. As a geodetic model
in analytical mechanics, (\ref{eq:16}) is based on the following
metric tensor $\mathcal{G}$ on the manifold of all metric tensors
in $V$: 
\begin{equation}
\mathcal{G}=Jg^{ik}g^{jl}\textrm{d}g_{ij}\otimes\textrm{d}g_{kl}+
Kg^{ij}g^{kl}\textrm{d}g_{ij}\otimes\textrm{d}g_{kl}. \label{eq:19}
\end{equation}
This metric tensor is evidently non-Euclidean and the corresponding Riemannian
structure on ${\rm Sym}(V^{*}\otimes V^{*})$ (or rather on its
submanifold consisting of the positive metrics $g$) has a non-vanishing
curvature tensor. In analogy to (\ref{eq:7}), (\ref{eq:8}) one can
put $g$ into interaction with ``matter'', 
e.g., with the ``particle'' of
mass $m$, moving in $M$, then the total kinetic energy ($x^{i}$
denoting particle coordinates) is given by 
\begin{equation}
T=T[g]+T_{\rm matt}[g, x]=\frac{I}{2}g^{ik}g^{jl}
\frac{\textrm{d}g_{ij}}{\textrm{d}t}\frac{\textrm{d}g_{kl}}{\textrm{d}t}+
\frac{K}{2}g^{ij}g^{kl}\frac{\textrm{d}g_{ij}}{\textrm{d}t}
\frac{\textrm{d}g_{kl}}{\textrm{d}t}+\frac{m}{2}g_{ij}
\frac{\textrm{d}x^{i}}{\textrm{d}t}\frac{\textrm{d}x^{j}}{\textrm{d}t}. 
\label{eq:20}
\end{equation}

Obviously, the same may be easily done for the system of particles. 
Expression (\ref{eq:20}) is based on the following metric tensor
$\mathcal{G}$ on $M\times{\rm Sym}^{+}(V^{*}\otimes V^{*})$:
\begin{equation}
\mathcal{G}=mg_{ij}\textrm{d}x^{i}\otimes\textrm{d}x^{j}
+Jg^{ik}g^{jl}\textrm{d}g_{ij}\otimes\textrm{d}g_{kl}
+Kg^{ij}g^{kl}\textrm{d}g_{ij}\otimes\textrm{d}g_{kl}. \label{eq:21}
\end{equation}
The analogy is obvious. The ``matter''
term may be extended by introducing some ``potential
energy'' $V$, e.g., as a function of the ``radial''
invariant 
\begin{equation}
r^{2}=g_{ij}x^{i}x^{j}. \label{eq:22}
\end{equation}

Kinetic energy (\ref{eq:20}) and its underlying metric (\ref{eq:21})
are invariant under the affine group ${\rm GAff}(M)$, which, after the choice
of some origin $\mathcal{O}\in M$, may be identified with the semi-direct
product ${\rm GL}(V)\underset{\sim}{\times}V$. The algebraic invariant
(\ref{eq:22}) is invariant only under ${\rm GL}(V)$, or more precisely, 
under the centro-affine group ${\rm GAff}(M, \mathcal{O})\subset {\rm GAff}(M)$
preserving the origin. When dealing with multi-particle material system
in $M$, one can obtain the total affine symmetry, replacing the quantity
(\ref{eq:22}) by the system of translationally-invariant functions
$r_{AB}$ on the configuration space $Q=M\times{\rm Sym}{}^{+}(V^{*}\otimes V^{*})$; these ``radial-like'' quantities are defined as 
\begin{equation}
r_{AB}=\sqrt{g_{ij}\left(x^{i}{}_{A}-x_{B}^{i}\right)
\left(x^{j}{}_{A}-x_{B}^{j}\right)}, \label{eq:23}
\end{equation}
where $x^{i}{}_{A}$ are affine coordinates of the $A$-th particle. 
It is seen that the symmetry ${\rm GAff}(M)$ is analogous to the general
covariance ${\rm Diff}\; M$ of General Relativity and those are just those
huge symmetry groups which are ``responsible''
for the strong, non-perturbative nonlinearities of the models. The
group ${\rm GAff}(M)$ acts on metrical degrees of freedom through its quotient
linear group ${\rm GL}(V)$, in the sense 
\begin{equation}
A\in {\rm GL}(V):\qquad[g_{ij}]\mapsto
\left[\left(A_{*}g\right)_{ij}\right]=
\left[g_{kl}\left. A^{-1}\right. ^{k}{}_{i}
\left. A^{-1}\right. ^{l}{}_{j}\right]. \label{eq:24}
\end{equation}

Let us mention, there are also some other generalized potentials, 
i.e., ones depending also on generalized velocities, compatible with
those invariance demands. But there is no place here for analyzing
this problem in more detail. 

Let us now concentrate on something else. In the toy models discussed
above we were dealing with models in analytical mechanics which had
two kinds of degrees of freedom: translational ones in the physical
affine space and internal ones, represented by the metric tensor
as a kind of collective variable. This picture remains with certain
instructive analogy with the metrical (generally-relativistic) model of
gravitation. Analytically, the internal modes were described by matrices. But we mentioned also about some other
class of mechanical models, namely, one where internal/collective degrees
of freedom were represented by elements of some Lie groups, to be
more precise, by elements of linear Lie groups, or by other manifolds
of linear mappings. Analytically they are also represented by matrices. 
But, and this is geometrically important, they are other objects, 
namely, mixed tensors, whereas scalar products are twice covariant
tensors. Objects with spaces of mixed tensors as configuration spaces
were investigated, e.g., by us and others in mechanics of so-called
affinely-rigid bodies. And they may be considered as mechanical toy
models of certain alternative models of gravitation, namely, tetrad
models in their various versions. And this has again very much, just
even more, to do with the link between symmetry and nonlinearity, 
and with some more or less diffused interpenetration between field theory and condensed matter physics, in particular, relativistic mechanics of structured continua. 

Let us again go back to gravitation ideas and relativistic structured media, 
this time within the framework of more or less modified tetrad models \cite{22,26,jjs_07}. Historically the tetrad models, originating from Weyl and Einstein, were thought on as some bridge between specially- and
generally-relativistic theories. Later on, it turned out that they
provide much richer class of models than the material Hilbert/Einstein
model. And the most deciding point was Dirac theory of generally relativistic
spinors, impossible to be formulated within the metrical framework. 
The main reason is that the universal covering group of ${\rm GL}(4,\mathbb{R})$
unlike ${\rm GL}(4,\mathbb{R})$ itself, is not linear, i.e., cannot be
faithfully realized by finite matrices. The same is true for any dimension
$n\geqslant3$. 

Instead of the metric field $g$, as the system of gravitational potential one uses the field of linear frames
$e=(\ldots, e_{A}, \ldots)$. 
Equivalently, one can use the dual field of co-frames $e^{-1}=(\ldots, e^{A}, \ldots)$, 
where 
\begin{equation}
\left\langle e^{A}, e_{B}\right\rangle =e^{A}{}_{\mu}e^{\mu}{}_{B}=\delta^{A}{}_{B};\label{eq:25}
\end{equation}
they uniquely determine each other. 

The next object is the corresponding teleparallelism connection $\Gamma[e,{\rm tel}]$ \cite{Kob-Nom63,22,jjs_07,Ster64}, defined uniquely by the condition that $e$ is parallel with respect
to the corresponding covariant differentiation $\nabla[e,{\rm tel}]$:
\begin{equation}
\nabla e_{A}=0, \qquad a=1, \ldots n. \label{eq:26}
\end{equation}
When no ambiguity occurs, we use the abbreviation $\Gamma[{\rm tel}]$, $\nabla[{\rm tel}]$, 
or just $\Gamma$, $\nabla$. One proves immediately that in local
coordinates
\begin{equation}
\Gamma[e,{\rm tel}]^{\mu}{}_{\nu\lambda}=e^{\mu}{}_{A}e^{A}{}_{\nu, \lambda}=
-e^{\mu}{}_{A, \lambda}e^{A}{}_{\nu}. \label{eq:27}
\end{equation}
Obviously, the curvature tensor of $\Gamma[tel]$ vanishes, but in
general its torsion 
\begin{equation}
S[e,{\rm tel}]^{\mu}{}_{\nu\lambda}=\frac{1}{2}e^{\mu}{}_{A}
\left(e^{A}{}_{\nu, \lambda}-e^{A}{}_{\lambda, \nu}\right)\label{eq:28}
\end{equation}
is a non-vanishing tensor; incidentally, it is familiar from the theory
of dislocations. Here it plays the role of tensorially invariant derivative
of the field $e$. 

Let $\eta$ denote some symmetric non-degenerate metric tensor in
the target space of the field $e$. Analytically it is given by a
constant and nonsingular symmetric matrix $[\eta_{AB}]$. For physical
reasons its signature is normal-hyperbolic; in the physical dimension
$n=4$, we usually put it as 
\begin{equation}
\left[\eta_{AB}\right]=\textrm{Diag}\left(1, -1, -1, -1\right). \label{eq:29}
\end{equation}
The contravariant inverse is denoted by $\left[\eta^{AB}\right]$:
\begin{equation}
\eta_{AC}\eta^{CB}=\delta_{A}{}^{B}. \label{eq:30}
\end{equation}

The Weyl-Dirac-Einstein metric tensor is given by 
\begin{equation}
h[e]=\eta_{AB}e^{A}\otimes e^{B}, \qquad h[e]_{\mu\nu}=\eta_{AB}e^{A}{}_{\mu}\otimes e^{B}{}_{\nu}.\label{eq:31}
\end{equation}

The linear group ${\rm GL}(n,\mathbb{R})$ (${\rm GL}(4,\mathbb{R})$ physically), 
as a structural group of the principal bundle of linear frames $FM$
or its dual $F^{*}M$, physically is a group of internal transformations
of the field $e$ (or dually, $e^{-1}$). It acts according to the
rule:
\begin{eqnarray}
e =  \left(\ldots, e_{A}, \ldots\right)&\mapsto& eL=\left(\ldots, e_{B}L^{B}{}_{A}, \ldots\right),\nonumber \\
\label{eq:32}\\
e^{-1}  =  \left(\ldots, e^{A}, \ldots\right)&\mapsto&\left(eL\right)^{-1}=
\left(\ldots, \left. L^{-1}\right. ^{A}{}_{B}e^{B}, \ldots\right), \nonumber \end{eqnarray}
for arbitrary $L\in {\rm GL}(n,\mathbb{R})$. This is a global action, 
in geometry one considers also the local action of the fields $L:M\rightarrow {\rm GL}(n,\mathbb{R})$, according to the rule:
\begin{equation}
e(x)\mapsto e(x)L(x), \qquad e(x)^{-1}\mapsto\left(e(x)L(x)\right)^{-1}.\label{eq:33}
\end{equation}
One can also restrict the values of $L$ to the Lorentz subgroup
${\rm O}(n,\eta)\subset {\rm GL}(n,\mathbb{R})$ (physically 
${\rm O}(1,3)\subset {\rm GL}(4,\mathbb{R})$), when 
\begin{equation}
\eta_{CD}L^{C}{}_{A}L^{D}{}_{B}=\eta_{AB}. \label{eq:34}
\end{equation}
Obviously, $h[e]$ is invariant under the local Lorentz action (\ref{eq:33}), (\ref{eq:34}). Unlike this, $S[e]$ is invariant only under the global
($x$-independent) action (\ref{eq:32}) of the total ${\rm GL}(n,\mathbb{R})$
and its Lorentz subgroup. But the both prescription $e\mapsto h[e]$ and 
$e\mapsto S[e]$ are generally covariant, 
\begin{equation}
h[\varphi_{*}e]=\varphi_{*}h[e], \qquad S[\varphi_{*}e]=\varphi_{*}S[e]. \label{eq:35}
\end{equation}
Using the tensors $h[e]$, $S[e]$ as algebraic brick-stones, one
can construct some byproduct quantities. First of all, let us quote
some tensors built of the tensor $S[e]$ alone:
\begin{eqnarray}
\gamma_{\mu\nu}&=&4S^{\alpha}{}_{\mu\beta}S^{\beta}{}_{\nu\alpha}  =\gamma_{\nu\mu},\qquad \gamma_{\mu}=S^{\alpha}{}_{\mu\alpha},\nonumber \\
\label{eq:36}\\
\Gamma_{\mu\nu}&=&4S^{\alpha}{}_{\beta\alpha}S^{\beta}{}_{\mu\nu} =-\Gamma_{\nu\mu}=2\gamma_{\alpha}S^{\alpha}{}_{\mu\nu}.\nonumber 
\end{eqnarray}
These are the only tensors built algebraically of $S$ alone in a quadratic
or linear ($\gamma_{\mu}$) way. 

Another important quantities are
scalars built in a quadratic way of $S$, with coefficients built algebraically
of $h[e]$, so-called Weitzenb\"{o}ck invariants \cite{22,26,jjs_07}:
\begin{eqnarray}
J_{1} & = & h_{\alpha\mu}h^{\beta\nu}h^{\gamma\varkappa}S^{\alpha}
{}_{\beta\gamma}S^{\mu}{}_{\nu\varkappa}, \nonumber \\
J_{2} & = & \frac{1}{4}h^{\mu\nu}\gamma_{\mu\nu}=h^{\mu\nu}S^{\alpha}
{}_{\mu\beta}S^{\beta}{}_{\alpha\nu}, \label{eq:37}\\
J_{3} & = & \frac{1}{4}h^{\mu\nu}\gamma_{\mu}\gamma_{\nu}=h^{\mu\nu}S^{\alpha}
{}_{\mu\beta}S^{\beta}{}_{\nu\alpha}. \nonumber 
\end{eqnarray}

There is also plenty of other concomitants of $S[e]$, $h[e]$, but
the above ones are distinguished by their property of be quadratic
in derivatives of $e$ (with the exception of $\gamma_{\mu}$, which
is linear in derivatives). No doubt, such quantities are geometrically
distinguished when constructing Lagrangians. Let us notice the particular
role of $\gamma_{\mu\nu}$. Being symmetric, it is an alternative
candidate for the metric tensor of $M$; alternative with respect
to the Einstein-Weyl-Dirac metric $h[e]$. It is the more important
that if $\left(\ldots, e_{A}, \ldots\right)$ form a semi-simple Lie
algebra with respect to the Lie bracket, 
\begin{equation}
\left[e_{A}, e_{B}\right]=C^{K}{}_{AB}e_{K}, \label{eq:38}
\end{equation}
 $C^{K}{}_{AB}$ being (structure) constants such that 
\begin{equation}
\det\left[C^{K}{}_{AL}C^{L}{}_{BK}\right]\neq 0, \label{eq:39}
\end{equation}
then locally $M$ may be identified with a semi-simple Lie group, 
and $\gamma[e]_{\mu\nu}$ becomes its non-degenerate Killing tensor \cite{Kob-Nom63,Ster64}. 
This is interesting, the more so that the signature of $\gamma_{\mu\nu}$
is not introduced ``by hand'';
instead it is a consequence of something more fundamental. 

It is important that all the above quantities are built of the field
$e$ in a generally-covariant way, so they satisfy 
\begin{equation}
F\left[\varphi_{*}e\right]=\varphi_{*}F\left[e\right], \qquad 
\varphi\in {\rm Diff}\; M. \label{eq:40}
\end{equation}
Their invariance status under internal transformations ${\rm GL}\left(n,\mathbb{R}\right)$, 
operating on the capital indices (target space transformations) is
a more complicated matter. Certainly the torsion tensor $S$, the
Killing and similar objects like (\ref{eq:36}), built algebraically
of $S$ alone, are invariant under the global action of 
${\rm GL}\left(n,\mathbb{R}\right)$, (\ref{eq:32}). 
The same concerns, of course, all ${\rm Diff}\; M$-invariant scalars built
of $S$ alone. Incidentally, one can show that all such scalars are
homogeneous functions of degree zero built of $S$. The dependence
of the Dirac-Weyl-Einstein metric $h[e]$ of $e$ is invariant under
the local action of ${\rm O}\left(1,3\right)\subset {\rm GL}\left(4,\mathbb{R}\right)$
(${\rm O}\left(n,\eta\right)\subset {\rm GL}\left(n,\mathbb{R}\right)$), i.e., 
under (\ref{eq:33}) with values of $L$ restricted to the Lorentz
subgroup. Certainly $e\mapsto h[e]$ is not invariant under ${\rm GL}\left(4,\mathbb{R}\right)$ (${\rm GL}\left(n,\mathbb{R}\right)$) even in the global sense. Moreover, it is not invariant under any subgroup of ${\rm GL}\left(4,\mathbb{R}\right)$
larger than the Lorentz group. The same concerns the Weitzenb\"{o}ck
invariants (\ref{eq:37}). There is however some delicate point with
very important and far-reaching consequences. Namely, substituting
to the Hilbert Lagrangian (\ref{eq:3}) the metric $h[e]$ instead of
$g$, one obtains the following expression for $\mathcal{L}_{\rm H}$ as
a function of $\left(e, \partial e\right)$:
\begin{equation}\label{eq:41}
\mathcal{L}_{\rm H}[e]=  -  \frac{1}{2\varkappa}\left(J_{1}+
2J_{2}-4J_{3}\right)\sqrt{\left|h[e]\right|}
 -  \frac{2}{\varkappa}\partial_{\mu}\left(S^{\alpha}
{}_{\alpha\beta}h^{\beta\mu}\sqrt{\left|h\right|}\right).
\end{equation}
The fourth term is a well-defined scalar density of weight one, because
it is a usual partial-derivative divergence of the contravariant vector
density of weight one. The symbol $\partial_{\mu}$ in this term may
be replaced by the covariant derivative $\nabla_{\mu}$ with respect
to the Levi-Civita affine connection built of $h[e]$. This fourth
term absorbs all second derivatives $\partial^{2}e$ of the basic
field $e$. Being a total divergence, it may be simply neglected. 
Then we obtain the effective Lagrangian explicitly free of second
derivatives, 
\begin{equation}
\mathcal{L}[e]=-\frac{1}{2\varkappa}
\left(J_{1}+2J_{2}-4J_{3}\right)
\sqrt{\left|h[e]\right|}.\label{eq:42}
\end{equation}
It is very important that unlike $G_{\rm H}\sqrt{\left|g\right|}$ in
(\ref{eq:5}), the expression (\ref{eq:42}) is a well-defined scalar
density of weight one. It is generally-covariant, 
\begin{equation}
\mathcal{L}[\varphi_{*}e]=\varphi_{*}\mathcal{L}[e],\qquad 
\varphi\in {\rm Diff}\; M. \label{eq:43}
\end{equation}

What concerns the total action of the internal Lorentz group, i.e., 
(\ref{eq:33}) with $L(x)$ satisfying (\ref{eq:34}) for any $x\in M$, 
obviously (\ref{eq:42}) is variationally invariant, i.e., invariant
modulo some divergence term. 

Once derived, (\ref{eq:42}) may be generalized to a wide class of
Lagrangians. First of all, the ratio of coefficients at $J$-s
needs not be necessarily $1:2:(-4)$. And in fact, it turns out that for a wide range of coefficients $c_{1}$, $c_{2}$, $c_{3}$ at $J_{1}$, $J_{2}$, $J_{3}$
such modified Lagrangians are compatible with experimental data. The
more serious modifications consist in admitting Lagrangians depending
in a general, including nonlinear, way of the Weitzenb\"{o}ck invariants, 
\begin{equation}
\mathcal{L}[e]=f(J_{1}, J_{2}, J_{3})
\sqrt{|h[e]|},\label{eq:44}
\end{equation}
$f$ being some real function of three variables. Nonlinearity of
such models may be incomparatively stronger than that of tetrad Hilbert
model (\ref{eq:42}) or its modified-coefficients version: 
\begin{equation}
\mathcal{L}[e]=(c_{1}J_{1}+c_{2}J_{2}
+c_{3}J_{3})\sqrt{|h|}. \label{eq:45}
\end{equation}
Historically, there were some attempts to avoid certain difficulties
of General Relativity, by admitting the form (\ref{eq:44}) \cite{22}. 

Obviously, models (\ref{eq:44}) other than (\ref{eq:42}) are only
globally, no longer locally, invariant under the internal Lorentz
group. But when once admitting such models, one is immediately faced
with the very natural temptation: why not to try to construct generally-covariant
Lagrangians $\mathcal{L}[e]$ invariant under the total linear group
${\rm GL}(4,\mathbb{R})$ (perhaps ${\rm GL}(n,\mathbb{R})$ in $n$-dimensional
space-times)? This is a very natural idea, because a priori ${\rm GL}(4,\mathbb{R})$
(${\rm GL}(n,\mathbb{R})$), the structure group of the principal bundle
of frames, is the most natural internal group, while any restriction
of ${\rm GL}(n,\mathbb{R})$ to a subgroup seems to be non-motivated by
any ``first principles''. Any Lagrangian
invariant under ${\rm Diff}\; M$ and ${\rm GL}(n,\mathbb{R})$ must be built
algebraically of $S$ alone. One can show that it is always a homogeneous
function of degree $n=\dim M$ of the tensor $S$. And now, something
very interesting results. It turns out that Lagrangians of this type
have automatically a generalized Born-Infeld structure \cite{BB79,BI34,26,jjs_07}. The simplest
of them have the form: 
\begin{equation}
\mathcal{L}[e]=\sqrt{|\det[L_{\mu\nu}]|}, \label{eq:46}
\end{equation}
where $L_{\mu\nu}$, referred too as the Lagrangian tensor, is a
linear combination of tensors $\gamma_{\mu\nu}$, $\gamma_{\mu}\gamma_{\nu}$, 
and $\Gamma_{\mu\nu}$, cf. (\ref{eq:36}): 
\begin{equation}
L_{\mu\nu}=A\gamma_{\mu\nu}+B\gamma_{\mu}\gamma_{\nu}
+C\Gamma_{\mu\nu}, \label{eq:47}
\end{equation}
where $A, B, C$ are real constants. One can also admit a purely imaginary
$C$. Then the tensor $L_{\mu\nu}$ is hermitian, $L_{\mu\nu}=\overline{L_{\nu\mu}}$, 
and its determinant is real. There exist also more complicated models, 
where $A, B, C$ are scalar functions of $S$. One can also multiply
the total square root by some scalars built of $S$ alone. One can
show that generally-covariant scalars built algebraically of $S$
alone are always homogeneous functions of degree zero. Typical structure
of such scalars is as follows: If $\gamma_{\mu\nu}$ happens to be
non-degenerate, then it may be used to the raising of indices. Then
one can construct of (\ref{eq:36}) various mixed tensors, their traces, etc;
in a natural way scalars may be obtained on the basis of trace-taking. 
Introducing of such scalars into (\ref{eq:46}), (\ref{eq:47}) would
complicate the model drastically, but probably without serious chance
for obtaining something essentially new. 

There are some very interesting peculiar features of the model (\ref{eq:46}),
(\ref{eq:47}). Namely, it is clear that Lagrangian $\mathcal{L}$, 
i.e., integrand of the action functional, is a scalar $W$-density
of weight one. This is necessary if the action $I$ (\ref{eq:11})
is to be a scalar quantity, and obviously, physically it must be so. 
There is one, almost canonical prescription for obtaining such densities. 
Namely, scalar densities of weight two usually appear as determinants
of matrices of twice covariant tensors. Then the scalar $W$-densities
are obtained as square roots of the absolute values of those determinants. 
Another approach is to interpret Lagrangian as a differential $n$-form. 
Those are in principle equivalent formulations, there are, however, 
some subtle points concerning orientation of $M$. 

In any case, it is quite natural to say that the primary notation
of variational theory is Lagrange tensor $L_{\mu\nu}$ depending algebraically
of a given field $\Psi$, its derivatives $\partial\Psi$ and eventually
also of the space-time point $x$ explicit, 
\begin{equation}
L_{\mu\nu}\left(x, \Psi(x), \partial\Psi(x)\right). \label{eq:48}
\end{equation}

What is usually referred to as Lagrangian $\mathcal{L}$, is a by-product, 
scalar density of weight one, given by (\ref{eq:46}). In commonly
used theories, one uses some metric tensor $g_{\mu\nu}$ and the scalar
representation $L$ of $\mathcal{L}$ given by the factorization:
\begin{equation}
\mathcal{L}=L\sqrt{\left|g\right|}. \label{eq:49}
\end{equation}
This is the language used in General Relativity. 

The ``Born-Infeld'' property of (\ref{eq:46}) is that it is the square-root of the determinant of ``something'', and the ``something'', i.e., Lagrange tensor, is a low-order, in this case second-order polynomial of field derivatives. 

In traditional Born-Infeld electrodynamics, Lagrangian has the form \cite{BB79,BI34}:
\begin{equation}
\mathcal{L}=b^{2}\sqrt{\left|\det\left[g_{\mu\nu}\right]\right|}
-\sqrt{\left|\det\left[bg_{\mu\nu}+F_{\mu\nu}\right]\right|}, \label{eq:50}
\end{equation}
where, obviously, the first term is non-dynamical. In Special Relativity
it is simply constant; in General Relativity it is $x$-dependent, 
but still independent on the electromagnetic field. Dynamical quantities
of the theory are $A_{\mu}$, i.e., components of the four-potential
covector; $F_{\mu\nu}$ is the electromagnetic tensor, 
\begin{equation}
F_{\mu\nu}=\partial_{\mu}A_{\nu}-\partial_{\nu}A_{\mu}. 
\label{eq:51}
\end{equation}

The only, rather artificial, role of the first term in (\ref{eq:50})
is to make the Lagrangian and energy vanishing when $F$ does vanish. 
The true dynamics is encoded purely in the second term. The Lagrange
tensor is given by
\begin{equation}
L_{\mu\nu}=bg_{\mu\nu}+F_{\mu\nu}, \label{eq:52}
\end{equation}
so, it is a first-order polynomial of derivatives of dynamical fields. 
It is just the peculiarity of the electromagnetic field that linear (or rather affine, linear-inhomogeneous) Lagrange tensors do exist, 
although some alternative models with quadratic dependence on derivatives
may be also constructed. In general, the quadratic dependence on derivatives
is the simplest possibility. The model (\ref{eq:50}) was motivated
by certain difficulties of classical electrodynamics. Due to the square-root
structure, it predicts the saturation of electromagnetic field, in
analogy to the maximal velocity, i.e., velocity of light in relativistic
point mechanics. Because of this, the energy of point charges, i.e., electromagnetic mass, was finite. For many physical reasons, the model (\ref{eq:50}), (\ref{eq:52}) is a canonical nonlinearity compatible with electrodynamics \cite{BB79,20}. Incidentally, in spite of certain current views, classical electrodynamics is still full of mysteries. Born-Infeld nonlinearity has to do with many of them \cite{BPT2009, PBGT2008, PB2008}. 
Let us notice, however, that in (\ref{eq:46}), (\ref{eq:47}) the
``Born-Infeld'' structure follows from something
very fundamental, namely, from the invariance assumptions: The model
was to be generally-covariant, i.e., invariant under ${\rm Diff}\; M$, and
invariant under ${\rm GL}(n,\mathbb{R})$ (physically ${\rm GL}(4,\mathbb{R})$)
as internal symmetry group. The ``huge''
symmetry ${\rm Diff}\; M\times {\rm GL}(n,\mathbb{R})$ just implies the ``Born-Infeld''
nonlinearity, i.e., self-interaction, as the simplest possible model. 
Lagrange tensor (\ref{eq:47}) is quadratic in derivatives of field
variables. In a sense, this is another ``pole''
of physical ``simplicity'', alternative
to linearity. In linear (and quasilinear) models, Lagrangians are
quadratic in derivatives. In ``Born-Infeld''
models it is no longer Lagrangian, but Lagrange tensor that is quadratic
(sometimes linear) in derivatives. In these models one deals with
the essential, non-perturbative nonlinearity, i.e., essential, strong
self-interaction. This self-interaction is deeply based on geometry. 
And due to this geometric background and the underlying symmetry group, 
the resulting nonlinearity is not artificially complicated. What concerns
simplicity, it is as close to linear systems as possible. This kinship
is based on the alternative: 
\begin{center}
\emph{Quadratic Lagrange Tensor --- Quadratic Lagrangian. } 
\par\end{center}
In a sense, in tetrad models, the opposition between field theory
and continuum theory (condensed matter) diffuses. From some point
of view, the tetrad field is a gravitational potential, but at the
same time it may be physically interpreted as the relativistic micromorphic
continuum. Roughly speaking, integral curves of the time-like ``legs''
of tetrads are world-lines of continuum particles. The remaining ``legs''
represent internal degrees of freedom of this continuum, attached
frames. So, it is really something like micropolar (Cosserat) or micromorphic
(Eringen) continuum medium \cite{Erin62,Erin68,Rub85,Rub86}. 

Now, let us close the circle of analogies in our study of essential
nonlinearities and geometric self-interactions. From General Relativity
we passed to its finite-dimensional models based on (\ref{eq:16}), 
(\ref{eq:17}), (\ref{eq:19})--(\ref{eq:21}), etc., 
where the ``spatial'' metric $g$ was a
kind of the internal/collective variable. Then, the tetrad models
of gravitation were briefly discussed, with the special stress on
the Born-Infeld type of self-interaction. Let us go back to finite-dimensional
analytical mechanics. There is an analogy between transition from General
Relativity to tetrad models and the transition from (\ref{eq:16}), 
(\ref{eq:17}), (\ref{eq:19})--(\ref{eq:21}) to
so-called affinely-rigid bodies, i.e., bodies rigid in the sense of
affine geometry \cite{Bur96,Cap89,Cap-Mar03,Gol01,Gol02,Gol03,Gol04,Gol04S,JJS74_2,JJS75_1,JJS75_2,JJS82_1,JJS82_2,1,JJS02_1,JJS02_2,JJS03,JJS04S,3,10,JJS-VK03,JJS-VK04,JJS-VK04S,all-book04,all04,all05,Dias94}. The configuration space of internal/collective modes
will be given then by ${\rm F}(V)$, the manifold of linear frames in $V$. 
So, the configuration space $M\times {\rm Sym}^{+}(V^{*}\otimes V^{*})$
will be replaced by $M\times {\rm F}(V)$. Generalized coordinates are $(x^{i}, e^{i}{}_{A})$, where $e^{i}{}_{A}$
are components of the frame vectors $e_{A}$ with respect to spatial
coordinates $x^{i}$. In analogy to (\ref{eq:31}) one can use the
metric 
\begin{equation}
h[e]=\delta_{AB}e^{A}\otimes e^{B}, \qquad h[e]_{ij}=\delta_{AB}e_{i}^{A}e_{j}^{B}, \label{eq:53a}
\end{equation}
where, obviously, $(\ldots, e^{A}, \ldots)$ is the co-frame dual to
$(\ldots, e_{A}, \ldots)$. If some metric tensor $g\in V^{\ast}\otimes V^{\ast}$
is fixed, then the usual reasoning leads to the following kinetic energy form:
\begin{equation}\label{eq:53b}
T=\frac{m}{2}g_{ij}\frac{\mathrm{d}x^{i}}{\mathrm{d}t}
\frac{\mathrm{d}x^{j}}{\mathrm{d}t}
+\frac{1}{2}g_{ij}\frac{\mathrm{d}e^{i}{}_{A}}{\mathrm{d}t}
\frac{\mathrm{d}e^{j}{}_{B}}{\mathrm{d}t}J^{AB}, 
\end{equation}
where $J^{AB}$ is the co-moving, thus, constant, tensor of inertia, i.e., quadrupole momentum of the mass distribution in representation of co-moving axes given by the moving frame $e=(\ldots, e_{A}, \ldots)$ \cite{Gol01, Gol02, Gol03, Gol04, Gol04S, Mart02, Mart03, Mart04_1, Mart04_2, Roz05, JJS74_2, 
JJS75_1, JJS75_2, JJS82_1, JJS82_2, 1, 26, JJS02_1, JJS02_2, JJS03, JJS04S, 3, 10, jjs_07, 
JJS-VK03, JJS-VK04, JJS-VK04S, all-book04, all04, all05}. If the mass distribution within the body is isotropic with respect to the co-moving frame, then 
\begin{equation}\label{eq:54}
J^{AB}=I\delta^{AB};
\end{equation}
this is an affine analogue of the spherical rigid body. 

Kinetic energy (\ref{eq:53b}) is invariant under the Euclidean group ${\rm E}(M, g)$, in particular, under spatial translations and under orthogonal group ${\rm O}(V,g)\subset {\rm GL}(V)$; more precisely, under its centro-affine versions ${\rm E}(M, g;\mathcal{O})$, where $\mathcal{O}\in M$ is a fixed origin in $M$. It is also invariant under the internal (material) $J$-orthogonal group ${\rm O}(n,J)\subset {\rm GL}(n,\mathbb{R})$. The latter group consists of matrices $L$ such that
\begin{equation}\label{eq:55}
L^{A}{}_{C}L^{B}{}_{D}J^{CD}=J^{AB}. 
\end{equation}
If $J$ is isotropic, then (\ref{eq:54}) holds and ${\rm O}(n,J)$ becomes just the usual orthogonal group ${\rm O}(n, \mathbb{R})$. 

In formulas (\ref{eq:16}), (\ref{eq:17}) we just objected against the fixed $g$; it was to be dynamical. If we follow the ideas of tetrad theory of gravitation, then it seems natural to do something else, namely, to substitute in (\ref{eq:53b}) $h[e]$ instead of $g$, 
\begin{equation}\label{eq:56}
T=\frac{m}{2}h[e]_{ij}\frac{\mathrm{d}x^{i}}{\mathrm{d}t}
\frac{\mathrm{d}x^{j}}{\mathrm{d}t}
+\frac{1}{2}h[e]_{ij}\frac{\mathrm{d}e^{i}{}_{A}}{\mathrm{d}t}
\frac{\mathrm{d}e^{j}{}_{B}}{\mathrm{d}t}J^{AB}. 
\end{equation}
This may be written as follows:
\begin{equation}\label{eq:57}
T=\frac{m}{2}\delta_{AB}\widehat{v}^{A}\widehat{v}^{B}
+\frac{1}{2}\delta_{AB}\widehat{\Omega}^{A}{}_{C}\widehat{\Omega}^{B}{}_{D}J^{CD}, 
\end{equation}
where $\widehat{v}^{A}$ and $\widehat{\Omega}^{K}{}_{L}$ are $e$-co-moving components of translational velocity and the so-called affine velocity, respectively:
\begin{equation}\label{eq:58}
\widehat{v}^{A}=e^{A}{}_{i}\frac{\mathrm{d}x^{i}}{\mathrm{d}t}=e^{A}{}_{i}v^{i}, \qquad  \widehat{\Omega}^{K}{}_{L}=e^{K}{}_{i}\frac{\mathrm{d}e^{i}{}_{L}}{\mathrm{d}t}. 
\end{equation}
The corresponding spatial affine velocity $\Omega^{i}{}_{j}$ is defined and related to $\widehat{\Omega}$ as follows:
\begin{equation}\label{eq:59}
\Omega^{i}{}_{j}=\frac{\mathrm{d}e^{i}{}_{K}}{\mathrm{d}t}
e^{K}{}_{j}=e^{i}{}_{K}\widehat{\Omega}^{K}{}_{L}e^{L}{}_{j}. 
\end{equation}
$\Omega$, $\widehat{\Omega}$ are Lie-algebraic objects, affine counterparts of angular velocity. Let us notice that $h[e]$ is identical with the Cauchy deformation tensor of elasticity theory, or rather, its special case corresponding to homogeneous (affine) deformations. 

It is very interesting that (\ref{eq:56}), (\ref{eq:57}) is invariant under ${\rm GAff}(M)$, the total affine group in $M$. This is a finite-dimensional counterpart of ${\rm Diff}\; M$, the group of general covariance in (\ref{eq:3}), (\ref{eq:42}), (\ref{eq:44}), (\ref{eq:45}). And if $J^{AB}=I\delta^{AB}$ (isotropy of the inertial tensor), then (\ref{eq:57}) becomes 
\begin{equation}\label{eq:60}
T=T_{\rm tr}+T_{\rm int}=
\frac{m}{2}\delta_{AB}\widehat{v}^{A}\widehat{v}^{B}
+\frac{I}{2}\delta_{AB}\widehat{\Omega}^{A}{}_{C}
\widehat{\Omega}^{B}{}_{D}\delta^{CD}
\end{equation}
and in addition to the spatial ${\rm GAff}(M)$-invariance, we have the ``internal'' invariance under ${\rm O}(n, \mathbb{R})$, the orthogonal group in $n$ dimensions. This is an analogue of the global internal Lorentz invariance in (\ref{eq:44}), (\ref{eq:45}). Physically, the ${\rm F}(V)$-degrees of freedom are collective/internal variables, which are in a sense more ``subtle'' than ${\rm Sym}^{+}(V^{*}\otimes V^{*})$. They introduce affine invariance and the corresponding essential nonlinearity. The question arises as to the finite-dimensional analogues of something like (\ref{eq:46}), (\ref{eq:47}). In other words: How to extend the internal orthogonal symmetry ${\rm O}(n, \mathbb{R})$ of (\ref{eq:60}) to the full linear group ${\rm GL}(n,\mathbb{R})$, in analogy to extending the Lorentz internal symmetry of (\ref{eq:44}), (\ref{eq:45}) to the full linear one like in (\ref{eq:46}), (\ref{eq:47})? To be honest, this is impossible for the total kinetic energy of affine body. It may be affinely invariant either in $M$ or internally in $\mathbb{R}^{n}$, but not simultaneously in both spaces. But the internal part may be affinely invariant both in $M$ and in $\mathbb{R}^{n}$; roughly speaking, simultaneously left  and right affinely invariant. 

And such models were already mentioned briefly after the formula (\ref{eq:16}). Namely, they are built of the second and first Casimir invariants, 
\begin{equation}\label{eq:61}
T[e]=\frac{I}{2}{\rm Tr}\left(\Omega^{2}\right)+\frac{K}{2}\left({\rm Tr}\; \Omega\right)^{2}=\frac{I}{2}{\rm Tr}\left(\widehat{\Omega^{2}}\right)
+\frac{K}{2}\left({\rm Tr}\; \widehat{\Omega}\right)^{2}. 
\end{equation}
The corresponding metric tensor on ${\rm F}(V)$ is given by 
\begin{equation}\label{eq:62}
\mathcal{G}=I\omega^{i}{}_{j}\otimes \omega^{j}{}_{i}+K\omega^{i}{}_{i}\otimes \omega^{j}{}_{j}=I\widehat{\omega}^{A}{}_{B}\otimes \widehat{\omega}^{B}{}_{A}+K\widehat{\omega}^{A}{}_{A}\otimes \widehat{\omega}^{B}{}_{B}, 
\end{equation}
where $I, K$ are constants and $\omega^{i}{}_{j}$, $\widehat{\omega}^{A}{}_{B}$ are differential forms on ${\rm F}(V)$ given respectively by the following formulas:
\begin{equation}\label{eq:63}
\omega^{i}{}_{j}=e^{A}{}_{j}\mathrm{d}e^{i}{}_{A}, \qquad \widehat{\omega}^{A}{}_{B}=e^{A}{}_{i}\mathrm{d}e^{i}{}_{B}, 
\end{equation}
thus, they are interrelated by
\begin{equation}\label{eq:64}
\omega^{i}{}_{j}=e^{i}{}_{A}e^{B}{}_{j}\widehat{\omega}^{A}{}_{B}. 
\end{equation}
Obviously, the main term is that controlled by $I$; the $K$-term is a merely correction, just in a complete analogy to (\ref{eq:17}), (\ref{eq:18}). 

The metric tensor $\mathcal{G}$ (\ref{eq:62}) on the manifold of frames ${\rm F}(V)$ is essentially Riemannian, its curvature tensor is non-vanishing. At the same time, this metric has a large isometry group ${\rm GL}(V)\times {\rm GL}(n,\mathbb{R})$, or rather its quotient with respect to the non-effectiveness kernel
\begin{equation}\label{eq:65}
\{(\lambda Id_{V}, \lambda^{-1}I_{n}):\lambda\in\mathbb{R}, \lambda\neq 0 \}. 
\end{equation}
Obviously, in the last formula, $Id_{V}$ denotes the identity mapping in $V$, and $I_{n}$ is the $n\times n$ identity matrix. 

The large isometry group is, as usual, correlated with the essential nonlinearity, i.e., essential self-interaction of the geodetic problem based on (\ref{eq:61}), (\ref{eq:62}). This is a finite-dimensional pattern for the field-theoretic models (\ref{eq:46}), (\ref{eq:47}). Affine symmetry and the absence of any metrical background in $M$ or $V$ are their common structural features. 

It turns out that dynamical models based on  (\ref{eq:46}), (\ref{eq:47}) and (\ref{eq:61}) possess certain interesting solutions \cite{Gol01, Gol02, Gol03, Gol04, Gol04S, Mart02, Mart03, Mart04_1, Mart04_2, Roz05, JJS74_2, 
JJS75_1, JJS75_2, JJS82_1, JJS82_2, 1, 26, JJS02_1, JJS02_2, JJS03, JJS04S, 3, 10, jjs_07, 
JJS-VK03, JJS-VK04, JJS-VK04S, all-book04, all04, all05}. The ones for (\ref{eq:46}), (\ref{eq:47}) admit certain cosmological interpretation. They are also interesting from the point of view of relativistic structured continuum. Geodetic models on ${\rm F}(V)$ or ${\rm GL}(V)$ based on (\ref{eq:61}), (\ref{eq:62}) are applicable in nonlinear elasticity. This is particularly suggestive when ${\rm GL}(V)$ is constrained to ${\rm SL}(V)$ or, equivalently, when in ${\rm F}(V)$ we impose holonomic constraints according to which the volume of frames is preserved. In anholonomic language such constraints may be described by any of the two equivalent conditions:
\begin{equation}\label{eq:66}
{\rm Tr}\; \Omega=0, \qquad {\rm Tr}\; \widehat{\Omega}=0. 
\end{equation} 
It turns out that although the group ${\rm SL}(V)$ is non-compact, and so is its any orbit in ${\rm F}(V)$, the geodetic models based on (\ref{eq:61}) as a Lagrangian predict some open family of bounded solutions describing nonlinear elastic vibrations, even without any use of potential energy. Above some threshold there is an open set of non-bounded ``escaping'' solutions (``dissociation threshold''). Such a model is interpretable from the point of view of integrable one-dimensional latices. The ``lattice points'' on $\mathbb{R}$ appear as deformation invariants. Without incompressibility constraints (\ref{eq:66}) (isochoric motion) the ``volume'' of $e$ is either constant or behaves in a singular way, collapsing to the point or infinitely expanding. However, this effect may be stabilized by introducing some auxiliary potential depending only on $\det[e^{i}{}_{A}]$ and preventing both the collapse and decay. 

The usefulness of the geodetic (no-potential) model of elastic vibrations consists in that its analysis may be to some extent reduced to calculating matrix exponents \cite{Gol01, Gol02, Gol03, Gol04, Gol04S, Mart02, Mart03, Mart04_1, Mart04_2, Roz05, JJS74_2, 
JJS75_1, JJS75_2, JJS82_1, JJS82_2, 1, 26, JJS02_1, JJS02_2, JJS03, JJS04S, 3, 10, jjs_07, 
JJS-VK03, JJS-VK04, JJS-VK04S, all-book04, all04, all05}. 

\section{General covariance versus Born-Infeld nonlinearity}

Our idea here is that there exists some link between essential, non-perturbative nonlinearity and invariance under ``large'' symmetry groups. More precisely, nonlinearities following from invariance demands turn out to be physically most interesting. The general covariance, i.e., invariance under ${\rm Diff}\; M$, so fundamental for General Relativity, is the best known example. Nonlinearity of Euler equations for ideal fluids is intimately connected with the invariance under the group of all volume-preserving diffeomorphisms of ${\mathbb R}^{3}$ \cite{Arn78,Binz91}. The characteristic Born-Infeld-type nonlinearity of our ``tetrad'' models (\ref{eq:46}), (\ref{eq:47}) is implied by the joint demand of general covariance and the invariance under ${\rm GL}(4,{\mathbb R})$ (${\rm GL}(n,{\mathbb R})$), i.e., internal invariance. Both models have some finite-dimensional counterparts, namely, (\ref{eq:16})--(\ref{eq:21}) and (\ref{eq:56}), (\ref{eq:60}), (\ref{eq:61}). Analytical mechanics offers here some toy models of general covariance and internal symmetry. Incidentally, those toy models may be quite practically useful as description of some internal or collective degrees of freedom. 

Two main ideas did appear: general covariance and Born-Infeld structure. Apparently, they seem to be different things, in spite of some generally-relativis\-tic motivation for the electrodynamical Born-Infeld ideas. Nevertheless, the Born-Infeld structure of (\ref{eq:46}), (\ref{eq:47}) was just a direct consequence of the demand of invariance under ${\rm Diff}\; M \times {\rm GL}(n,{\mathbb R})$ (physically $n=4$). Is this accidental?

A more general question arises as to the very relationship between general covariance and the Born-Infeld type of nonlinearity, namely, the characteristic square-root structure and the second-order polynomial dependence of the Lagrange tensor $L_{\mu\nu}$ on field derivatives. In our model (\ref{eq:46}), (\ref{eq:47}) it was a quite canonical kinship. How is it in general?

Let us begin with the very idea of general covariance. Not every kind of physical field does admit a generally-covariant variational principle. The twice covariant tensor, e.g., metric tensor, does it. And every system of fields containing metric does so as well. This is the very idea of General Relativity. And there was a wrong view that it is the only possibility. Obviously, the tetrad ($n$-leg) field is also good from this point of view. But what are other possibilities?  The peculiarity of the metric field is that it is a twice covariant tensor field, i.e., field of scalar products. What about mixed second-order tensors, i.e., fields of linear mappings? Let $X$, analytically $X^{\mu}{}_{\nu}$ be such a field. It turns out that $X$ does admit a generally-covariant variational principle, but the simplest thing one can invent is rather complicated and the only possibility is just of the ``Born-Infeld'' type. Namely, it is a well-known fact that with every pair of mixed (once contravariant and once covariant) tensor fields $X$, $Y$ one can associate so-called Nijenhuis torsion $S(X, Y)$, which is once contravariant, twice covariant and antisymmetric in covariant indices. Perhaps it would be rather obscuring to quote the explicit formula, which belongs to the realm of advanced differential geometry \cite{Kob-Nom63}; in any case the point is that $S(X,Y)$, analytically $S(X, Y)^{\mu}{}_{\nu \lambda}$, is algebraically built of the components of $X$, $Y$ and their first-order derivatives. For any vector field $X$ we can invariantly define the tensor field $S(X):=S(X, X)$ and its byproducts, like, e.g., the Lagrange tensor
\begin{equation}
L[X]_{\mu \nu}=AS^{\lambda}{}_{\mu \varkappa}S^{\varkappa}{}_{\nu \lambda} + BS^{\lambda}{}_{\mu \lambda}S^{\varkappa}{}_{\nu \varkappa} + CS^{\lambda}{}_{\varkappa \lambda}S^{\varkappa}{}_{\mu \nu}, \label{eq:67}
\end{equation}
where $A$, $B$, $C$ are real constants. Nothing more natural (``more clever'') may be invented. The only possibility of generally-covariant Lagrangian is just (\ref{eq:46}) with (\ref{eq:67}) as the Lagrange tensor. 

It would be difficult to decide at this stage what would be the physical usefulness of such models. Nevertheless, they are well-defined and they witness that the Born-Infeld scheme in many situations is the only one compatible with geometry of degrees of freedom. 

But let us try to be more concrete with the problem. The question
is what might be a general scheme for generally-covariant field models. 
In this formulation it is too general to be effectively discussed. We know about models of degrees of freedom admitting generally-covariant
dynamical principles. Those are, among others, analytically speaking, 
``matrix fields'', like, e.g., the metric field, i.e., the field of twice covariant tensors, the field of mixed tensors, or the field of co-frames, i.e., analytically speaking, the $n$-tuple of covector fields on the $n$-dimensional
manifold (``space-time''). In any case it is so that in a generally-covariant
field theory in $n$-dimensional ``space-time''
manifold the field must have more than $n$ components. Because, 
roughly speaking, the general covariance may reduce any of $n$ field
components to an arbitrarily given function form, e.g., identifying
them locally with ``space-time'' coordinates. The simplest, academic
model is that of some $N$-component scalar field in $n$-dimensional
``space-time'' \cite{BB79,31,24}. As mentioned the general covariance implies that $N>n$, otherwise any generally-covariant model will be either trivial (every
field is a solution) or empty (no solutions at all). 

Let the target space $V$ of dimension $N$ be endowed with some pseudo-Riemannian
metric $\eta$. Obviously, the simplest situation is one when $V$
is a linear space and $\eta\in V^{*}\otimes V^{*}$ is some ``constant'' (pseudo-)Euclidean metric. Quite well we may assume $V$ to be
a complex linear space and $h$ some sesqulinear hermitian form. However, 
let us fix attention on the simplest case of a linear space with (pseudo-)Euclidean metric. Any $V$-valued scalar field $\Psi:\; M\rightarrow V$, 
i.e., analytically speaking, the multiplet of $N$ real scalar fields
$\Psi^{A}$ on $M$ (some basis in $V$ fixed) induces some kind of
$\Psi$-dependent metric in $M$, namely, the pull-back 
\begin{equation}
g[\Psi]=\Psi^{*}\cdot\eta, \qquad g[\Psi]_{\mu\nu}=\eta_{AB}\Psi^{A}{}, _{\mu}\Psi^{B}{}, _{\nu};\label{eq:68}
\end{equation}
comma, as usual, denoting the partial derivative. 

Let us stress that there
is no fixed metric in $M$, the space-time manifold is absolutely
amorphous. The only absolute element, $\eta\in V^{*}\otimes V^{*}$,
is an inhabitant of the target space $V$, it leaves in ``Heaven'', 
not in ``Earth''. Incidentally, there are linear spaces or manifolds
with intrinsic metrics, e.g., Lie algebras, Lie groups, manifolds of
scalar products as discussed above, etc. So, this ``absolutism'' of
$\eta$ need not be taken too seriously. Let us stress that in this
sense our tetrad/$n$-leg model was just completely amorphous, because
nothing but the linear space structure was assumed in the target space
$\mathbb{R}^{n}$ (physically $\mathbb{R}^{4}$). The simplest model
now is one, in which the Lagrange tensor $L_{\mu\nu}$ just coincides
with $g_{\mu\nu}$. One can also consider some ``potential'' terms
$U$, e.g., ones built of $\left\Vert \Psi\right\Vert ^{2}=\eta_{AB}\Psi^{A}\Psi^{B}$, 
and take 
\begin{equation}
L_{\mu\nu}=U\left(\Psi\right)g[\Psi]_{\mu\nu}. \label{eq:69}
\end{equation}
Of course, this is a re-definition of $\eta$ in a sense and it is
essential only when $\eta$ is a constant (pseudo-Euclidean) metric
in a linear space. 

Euler-Lagrange equations resulting from (\ref{eq:68}) with $U=1$ may
be invariantly written down as follows: 
\begin{equation}
g^{\mu\nu}\nabla_{\mu}\nabla_{\nu}\Psi^{A}=0, \qquad A=1, \ldots, N,
\label{eq:70}
\end{equation}
where, obviously, $g^{\mu\nu}$ are components of the contravariant
inverse of $g[\Psi]$, and $\nabla_{\mu}$ are operators of the covariant
differentiation in the sense of the Levi-Civita connection built of
$g$. Although (\ref{eq:70}) formally looks like the d'Alembert equation
for the multiplet $\Psi^{A}$, $A=1,\ldots N$, this system of differential
equations is strongly nonlinear, just essentially, non-perturbatively
nonlinear, because $g^{\mu\nu}$ and the Christoffel coefficients depend
on $\Psi$. This dependence results in the mutual coupling of equations. 
For any fixed $A$, the coefficients in the equation for $\Psi^{A}$ depend
on all the fields $\Psi^{B}$. 

The nice invariant form (\ref{eq:70}) may be explicitly, but ugly, 
written down as follows: 
\begin{equation}
g^{\mu\nu}\Psi^{A}{}_{, \mu\nu}+\Psi^{A}{}_{, \nu}\left(\frac{1}{2}
g^{\mu\nu}g^{\alpha\beta}-g^{\mu\alpha}g^{\nu\beta}\right)
g_{\alpha\beta, \mu}=0. \label{eq:71}
\end{equation}

Geometrical meaning of equations (\ref{eq:69}), (\ref{eq:70}) is
that the submanifold $\Psi(M)\subset V$ is a minimal surface in the
sense of pseudo-Euclidean geometry; its mean curvature does vanish \cite{31}. 
This geometrical interpretation is generally true, not only in the situation when $(V, \eta)$ is flat. The system (\ref{eq:70}) is redundant; this is a consequence of the gauge arbitrariness corresponding to the general covariance. The ${\rm Diff}\; M$-invariance implies that among the $N$ fields $\Psi^{A}$ there are, roughly speaking, only $(N-n)$ independent ones, while $n$ equations are superfluous and have the status of identities. There are $n$ purely gauge variables among $\Psi^{A}$ and those may be fixed by something like coordinate conditions. The simplest, although in a sense most ``brutal'', way of eliminating gauge variables is to identify some $n$-tuple of fields $\Psi^{A}$, e.g., $\Psi^{\mu}$, $\mu=1, \ldots, n$ (physically $n=4$), with space-time coordinates, i.e., to put 
\begin{equation}\label{eq:72}
\Psi^{\mu}=x^{\mu}, \qquad \mu=1, \dots, n. 
\end{equation}
The gauge condition may be chosen in the form:
\begin{equation}\label{eq:73}
\left( \frac{1}{2}g^{\mu\nu}g^{\alpha\beta}-g^{\mu\alpha}g^{\nu\beta}\right)g_{\alpha\beta, \mu}=0, 
\end{equation}
quite independently on the convention (\ref{eq:72}); this condition is more general. This gauge condition implies that (\ref{eq:71}) acquires the usual ``d'Alembert'' form in the sense that
\begin{equation}\label{eq:74}
g^{\mu\nu}\Psi^{A}{}_{,\mu\nu}=0. 
\end{equation}
Obviously, equation (\ref{eq:73}) is non-tensorial and this is correct, otherwise it would not be coordinate condition, i.e., fixation of gauge within the ${\rm Diff}\; M$-invariant scheme. If we assume $(\ref{eq:72})$, what locally is always correct (globally there are, obviously, some subtle problems), then our gauge equations (\ref{eq:73}) become identities, they are trivially satisfied. The fields $\Psi^{a}$, $a=n+1, \ldots, N$, are genuine degrees of freedom. 

It is convenient to choose coordinates in the target space in such
a way that the matrix of $\eta$ splits into blocks:
\begin{equation}
h_{\mu a}=0, \qquad\mu=1, \ldots, n, \qquad a=n+1, \ldots, N.\label{eq:75}
\end{equation}
Then we have 
\begin{equation}
g_{\mu\nu}=\eta_{\mu\nu}+\eta_{ab}\Psi^{a}{}_{, \mu}\Psi^{b}{}_{, \nu}, 
\label{eq:76}
\end{equation}
the summation convention is meant in the sense of indices $a, b=n+1, \ldots, N$. The true dynamics, free of gauge, is described by ``d'Alembert''
equations for $\Psi^{a}$:
\begin{equation}
g^{\mu\nu}\Psi^{a}{}_{, \mu\nu}=0, \qquad a=n+1, \ldots, N. \label{eq:77}
\end{equation}
These are Euler-Lagrange equations for the effective Lagrangian based
on the effective Lagrange tensor 
\begin{equation}
L\left({\rm eff}\right)_{\mu\nu}=\eta_{\mu\nu}+\eta_{ab}
\Psi^{a}{}_{, \mu}\Psi^{b}{}_{, \nu}. 
\label{eq:78}
\end{equation}
The coefficients $\eta_{\mu\nu}$ play the role of something like
the analytical representation of some fixed space-time metric, although, 
as a matter of fact, such a metric was not assumed here. Without the
block-structure assumption (\ref{eq:75}), the effective Lagrange tensor
would be given by 
\begin{equation}
L\left({\rm eff}\right)_{\mu\nu}=\eta_{\mu\nu}+2\eta_{a(\mu}\Psi^{a}
{}_{, \nu)}+\eta_{ab}\Psi^{a}{}_{, \mu}\Psi^{b}{}_{, \nu}, \label{eq:79}
\end{equation}
again with the summation convention extended over Latin indices 
$a=n+1, \ldots, N$. 

The Born-Infeld structure, in the classical form known from electrodynamics, is easily readable here. Under the square-root sign in the effective Lagrangian we recognize the field-independent effective metric $\eta_{\mu \nu}$, the term linear in (gauge-free) field derivatives, and the term quadratic in field derivatives. The linear one is like in Born-Infeld electrodynamics, the quadratic one resembles our tetrad model (\ref{eq:46}), (\ref{eq:47}). Nevertheless, let us mention that the term quadratic in $F_{\mu \nu}$ is possible also in certain modifications of Born-Infeld electrodynamics. The corresponding Lagrange tensor might be given by something like 
\begin{equation}
L_{\mu \nu}=bg_{\mu \nu}+ F_{\mu \nu} + dg^{\alpha \beta}F_{\mu \alpha}F_{\beta \nu} + kg^{\alpha \beta}g^{\varkappa \lambda} F_{\alpha \varkappa}F_{\beta \lambda} g_{\mu \nu}, \label{eq:80}
\end{equation}
where $b$,  $d$,  $k$ are constants. For fields which are not too strong, the predictions of (\ref{eq:52}), (\ref{eq:72}) are in good agreement. It is other thing that (\ref{eq:72}) certainly will not show some important features of the traditional model (\ref{eq:52}), because the latter one is canonical and unique in certain sense. 

The scalar Born-Infeld models based on the effective Lagrange tensors (\ref{eq:78}) with $N=n+1$ (physically $n=4$) was used in nonlinear scalar optics, i.e., in situations where the polarization effects may be neglected. In any case, we have found above the link between general covariance and the Born-Infeld structure of Lagrangians based on Lagrange tensors with at most quadratic dependence of field derivatives. In the scalar Born-Infeld electrodynamics solutions appear as stationary surfaces (``minimal surfaces'') of dimension four in the five-dimensional target space $V$. The corresponding metric $\eta$ has the signature $({+}{+}{-}{-}{-})$. If we denote
\begin{equation}
\left[\eta_{ab}\right] = {\rm diag} (\eta, 1, -1, -1, -1), \label{eq:81}
\end{equation}
then the effective Lagrangian is based on Lagrange tensor
\begin{equation}
L\left({\rm eff}\right)_{\mu \nu}= \eta_{\mu \nu}+ \eta \Psi_{, \mu}\Psi_{, \nu}. \label{eq:82}
\end{equation}

It is interesting that for the field $\Psi$ we obtain solutions of exactly the same form as one for the scalar potential $A_{0}$ in the ``usual'' four-covector electrodynamics:
\begin{equation}
\Psi(r)=\sqrt{\frac{A}{\eta}} \int^{r}_{0}\frac{\mathrm{d}x}{\sqrt{A+x^{4}}}, \label{eq:83}
\end{equation}
where $A>0$ is some integration constant. This small fact is very interesting in itself. 

To summarize, let us repeat some important special cases, which are not only suggestive but also physically interpretable. 
\begin{enumerate}
\item $N$ is arbitrary, $n=1$ --- geodetic curves.

\item $N=3$, $n=2$ --- rubber films, soap bubbles, etc. 

\item $N=4$, $n=1$, $\eta$ is Minkowskian. This is relativistic point mechanics. Obviously, for the free particle the effective Lagrangian is given by 
\begin{equation}
L\left({\rm eff}\right)=-mc^{2} \sqrt{1-\frac{v^{2}}{c^{2}}}. \label{eq:84}
\end{equation}

\item $N=4$, $n=2$, $\eta$ is Minkowskian. These are strings, 'tHooft-Polyakov-Kleinert models.

\item $N$ is arbitrary, $n=1$, $\eta$ is Riemannian, $U=2(E-V)$, cf. (\ref{eq:69}), $E$ is the fixed total energy, $V$ is potential. We easily recognize the Maupertuis variational principle. 
\end{enumerate}

The scalar models provide an interesting ``Kunst der Fuge'', the exercise for the study of essential nonlinearity in the context of general covariance. Let us stress that our $n$-leg model is in a sense ``better'' than all scalar Born-Infeld model, because it does not  assume any target metric. 

\section*{Acknowledgements}

We are very grateful to our friend professor Anatoly K. Prykarpatsky for our fruitful and inspiring discussions. Some of our results were obtained within the framework of the research project 501 018 32/1992 financed from the Scientific Research Support Fund in 2007-2010. We are greatly indebted to the Ministry of Science and Higher Education for this financial support. The support within the framework of Institute internal programme 203 is also greatly acknowledged.

\end{document}